\documentclass[conference]{IEEEtran}
\usepackage{amsmath,amsfonts}
\usepackage{booktabs}
\usepackage{algorithmic}
\usepackage{algorithm}
\usepackage{array}
\usepackage[caption=false,font=normalsize,textfont=sf]{subfig}
\usepackage{textcomp}
\usepackage{stfloats}
\usepackage{url}
\usepackage{verbatim}
\usepackage{graphicx}
\usepackage{cite}
\usepackage{multicol}
\usepackage{amssymb}
\usepackage{lineno}
\usepackage{makecell}
\usepackage{indentfirst}
\usepackage{bm}
\usepackage{color}
\usepackage{cite}
\usepackage{epstopdf}
\usepackage{amssymb}
\usepackage{xcolor}
\usepackage{tikz}
\usepackage{mathtools}
\usepackage{pifont}
\newcommand{\tr}{\mathrm{tr}}
\newcommand{\HH}{\mathrm{H}}	
\newcommand{\TT}{\mathrm{T}}

\newtheorem{proposition}{Proposition}

\newtheorem{lemma}{Lemma}
\begin{document}
	
	\title{Sensing Mutual Information with Random Signals in Gaussian Channels}

\author{Lei~Xie$^{\star}$, Fan Liu$^{\dagger}$, Zhanyuan Xie$^{\ddagger}$, Zheng Jiang$^{\ddagger}$, and Shenghui Song$^{\star}$\\
{$^{\star}$The Hong Kong University of Science and Technology, Hong Kong}\\
{$^{\dagger}$Southern University of Science and Technology, China}\\
{$^{\ddagger}$China Telecom Research Institute, China}
%{Email: $\{$eelxie, eeshsong$\}$@ust.hk}
}

	\maketitle
	
	\begin{abstract}
	Sensing performance is typically evaluated by classical metrics, such as Cramér-Rao bound and signal-to-clutter-plus-noise ratio. 
	The recent development of the integrated sensing and communication (ISAC) framework motivated the efforts to unify the metric for sensing and communication, where researchers have proposed to utilize mutual information (MI) to measure the sensing performance with deterministic signals. However, the need to communicate in ISAC systems necessitates the use of random signals for sensing applications and the closed-form evaluation for the sensing mutual information (SMI) with random signals is not yet available in the literature. 
	This paper investigates the achievable performance and precoder design for sensing applications with random signals. 
	For that purpose, we first derive the closed-form expression for the SMI with random signals by utilizing random matrix theory. The result reveals some interesting physical insights regarding the relation between the SMI with deterministic and random signals. 
	%For that purpose, we first propose a sensing performance metric called sensing MI (SMI) to measure the sensing performance with random signals. An explicit expression for the SMI is then derived by utilizing random matrix theory, revealing some interesting physical insights. 
	The derived SMI is then utilized to optimize the precoder by leveraging a manifold-based optimization approach. The effectiveness of the proposed methods is validated by simulation results. 
	\end{abstract}
	
	
	\section{Introduction}

	With the development of innovative applications that demand accurate environment information, e.g., autonomous driving and unmanned aerial vehicle (UAV) networks, sensing becomes an essential requirement for future wireless networks. To this end, the integrated sensing and communications (ISAC) framework has attracted much attention  \cite{liu2022integrated,xie2023collaborative}.  
	However, there are still substantial challenges that need to be tackled before we can fully unleash the potential of ISAC. For example, sensing and communication systems have been developed separately for many years with their own performance metrics. 
	In particular, sensing performance is normally characterized by metrics such as Cram\'{e}r-Rao bound (CRB) \cite{liu2021cramer}, signal-to-clutter-plus-noise ratio \cite{xie2022perceptive}, and minimum mean-square error \cite{herbert2017mmse}, while communication performance is usually measured by mutual information (MI) \cite{hachem2008new} and outage probability \cite{ko2000outage}. The inconsistency between the performance metrics of two systems poses challenges to integrative design. 
	
	%the conventional radar systems commonly employ well-designed deterministic signals for target sensing, but random signals are necessary for communication systems to convey information. 
	%While there have been studies focusing on the design of ISAC system with random signals \cite{lu2023sensing}, the sensing performance with random signals has not been well understood.
	%However, the sensing performance with random signals has not been well understood. 
	
	%Sensing performance are normally characterized by metrics such as Cramer-Rao Bound (CRB) \cite{liu2021cramer}, signal-to-clutter-plus-noise ratio (SCNR) \cite{xie2022perceptive}, minimum mean-square error (MMSE) \cite{herbert2017mmse}, detection probability \cite{xie2022networked}, and mutual information (MI) \cite{tang2018spectrally}. 
	%Among them, MI, which is widely utilized to measure communication performance, represents a promising metric to unify the metric for both sensing and communication. 

	Recently, some research efforts have been devoted to unifying the performance metric for sensing and communication. For that purpose, MI has been utilized to measure sensing performance \cite{bell1993information,tang2018spectrally,yang2007mimo,liu2023deterministic,dong2023rethinking}. For example, %the authors of \cite{bell1993information} demonstrated that the waveforms that maximize the MI between the random target response and the received signals lead to enhanced sensing performance. The  
	the authors of \cite{yang2007mimo} showed that the radar waveform that maximizes the MI between the Gaussian-distributed target response and the received signal also minimizes the MMSE for estimating the target response. %Therefore, to unify the metric for both communication and sensing, we employ the MI between the received signals and the target response matrix as the sensing metric. 
	However, existing works \cite{bell1993information,tang2018spectrally,yang2007mimo} mainly focused on the sensing MI (SMI) with deterministic signals, and the result is not applicable to ISAC systems that utilize random signals for joint sensing and communication purposes. Recently, some works considered the sensing performance with random signals and revealed the intrinsic connection between SMI and other metrics from a rate-distortion perspective \cite{liu2023deterministic,dong2023rethinking}. However, a closed-form expression for the SMI of systems utilizing random signals is not yet available.

	This paper has two objectives: 1) to evaluate the sensing performance with random signals; and 2) to optimize the sensing performance through signal design. 
	To achieve the first objective, we derive a closed-form expression for the SMI between the random target response and the received signals, where random signals are utilized. 
	%propose to utilize the SMI between the random target response and the measurement to measure the average sensing performance with random signals. Then, we derive an explicit expression for the SMI. 
	It is shown that the SMI with random signals is upper-bounded by that with a properly-designed deterministic signal. The performance loss due to the use of random signals is also investigated. 
	Building upon the theoretical result, we maximize the SMI by optimizing the transmit signal with manifold optimization. 
	Simulations results show that the proposed method achieves superior performance than existing methods. 

	The remainder of the paper is organized as follows.  
	Section II presents the system model. Section III provides a closed-form expression for the SMI and reveals some physical insights. 
	Section IV introduces a manifold-based approach to maximize the SMI by optimizing the transmitted precoder. 
	Simulation results are given in Section V to validate the effectiveness of the proposed manifold-based method.	
	Finally, Section V concludes the paper, summarizing the key findings and contributions.

	\section{System Model}
	Consider a bi-static sensing system, which is composed of a sensing transmitter equipped with $N_T$ antennas and a sensing receiver with $N_R$ antennas. Assume that the target sensing is performed within a coherent processing interval (CPI) consisting of $N_S$ frames. %, as illustrated in Fig. \ref{fig_sys}.
%	\begin{figure}[!t]
%		\centering
%		\includegraphics[width=3.0in]{PMN-TMT-2.pdf}
%		\caption{Illustration of the considered ISAC system. }
%		\label{fig_sys}
%	\end{figure}
%	
	The baseband received signal in a CPI is given by \cite{tang2018spectrally}
\begin{equation}
	\begin{split}
		\mathbf{Y} =   \mathbf{H} \mathbf{X}+ \mathbf{N} \in \mathcal{C}^{N_R\times N_S},
	\end{split}
\end{equation}
where $\mathbf{X}\in \mathcal{C}^{N_T\times N_S}$ denotes the transmitted signal, $\mathbf{H}$ represents the target response matrix, and $\mathbf{N} \in \mathcal{C}^{N_R\times N_S}$ is the addictive white Gaussian noise. The target response matrix can be given by 
\begin{equation}
	\mathbf{H} = \sum_{k=1}^{K} \epsilon_{k} \mathbf{b}(\vartheta_k) \mathbf{a}^\HH(\phi_{k})\in \mathcal{C}^{N_R\times N_T}
\end{equation}
where $K$ denotes the number of sensing targets. 
Note that $\epsilon_{k}$ represents the complex reflection coefficient of the $k$th target and its amplitude is determined by the path loss and the radar cross section (RCS). The reflection coefficient is normally assumed to follow Gaussian distribution with $\epsilon_{k}\sim \mathcal{CN}(0,\sigma_{k}^2)$. 
%Assume the complex coefficient of the $k$th target whose amplitude captures the path loss and the radar cross section, denoted as $\epsilon_{k}$, follows Gaussian distribution with  $\epsilon_{k}\sim \mathcal{CN}(0,\sigma_{k}^2)$. %In this paper, we adopt the Swerling I model where $\epsilon_{k}$ remains constant in a CPI and varies between CPIs. 
The two vectors, i.e., $\mathbf{a}(\cdot)$ and $\mathbf{b}(\cdot)$, denote the steering vectors for the transmit and receive array, respectively, and $\vartheta_k$ and $\phi_{k}$ represent the angle of departure (AOD) and the angle of arrival (AOA) for the $k$th target, respectively. %Consequently, $\mathbf{H}$ follows a Gaussian distribution, and varies between CPIs in an independent identically distributed (i.i.d.) manner.
%In one CPI, we assume that $\{\vartheta_k\}_{k=1}^K$ and $\{ \phi_{k}\}_{k=1}^K$ are fixed, and randomly varies every Ns symbols in an i.i.d. manner

Different from conventional radar systems which typically use deterministic signal, ISAC systems use random signals for communication purposes. In this paper, we consider the widely adopted Gaussian signal with $\mathbf{X} = \mathbf{F} \mathbf{S}$, where $\mathbf{F}\in \mathcal{C}^{N_T\times K}$ represents the precoder matrix and $\mathbf{S}\in \mathcal{C}^{K\times N_S}$ denotes a random matrix whose entries are independent and identically distributed (i.i.d.) Gaussian random variables with variance $\frac{1}{N_S}$ and $N_S \geq K$. It follows that  $\mathbb{E}(\mathbf{S}\mathbf{S}^\HH)=\mathbf{I}$. 
By stacking the columns of $\mathbf{Y}^\HH$, we can obtain
\begin{equation}
	\begin{split}
		\mathbf{y} \triangleq \text{vec} (\mathbf{Y}^\HH)= \mathring{\mathbf{X}}^\HH \mathbf{h} + \mathbf{n} \in \mathcal{C}^{N\times 1},
	\end{split}
\end{equation}
where $N = N_TN_R$, $\mathring{\mathbf{X}} = \mathbf{I}_{N_R} \otimes \mathbf{X} $, $\mathbf{h}= \text{vec}(\mathbf{H}^\HH)$, and $\mathbf{n}= \text{vec}(\mathbf{N}^\HH)\sim\mathcal{CN}(\mathbf{0},\sigma_N^2\mathbf{I}_N)$. 
The channel correlation matrix $\mathbf{R}\triangleq \mathbb{E}(\mathbf{h}\mathbf{h}^\HH)$ can be well approximated by $\mathbf{R} = \mathbf{R}_R \otimes \mathbf{R}_{T}$ \cite{965098,1021913,1300860}, where $\mathbf{R}_R$ and $\mathbf{R}_{T}$ denote the correlation matrices at the receiver and the transmitter, respectively, with $\mathrm{rank}(\mathbf{R}_R)\leq K$ and $\mathrm{rank}(\mathbf{R}_T)\leq K$. %Such a model has been widely applied to the MIMO radar systems \cite{965098,1021913,1300860}.
%As a result, 
In this paper, we assume $\mathbf{R}_R$ and $\mathbf{R}_{T}$ are available to the transmitter. This assumption is justified in ISAC systems, where the prior knowledge of the targets can be obtained from previous estimates \cite{herbert2017mmse,tang2018spectrally}.

\section{Sensing mutual information}
To detect the target, the sensing receiver will estimate the target response vector $\mathbf{h}$ based on the received signals $\mathbf{y}$, or $\mathbf{H}$ based on $\mathbf{Y}$, equivalently \cite{tang2018spectrally,ren2023fundamental}.
Assuming a deterministic sensing signal, previous works focused on optimizing $\mathbf{S}$ to maximize the SMI between the received signals and the target response \cite{tang2018spectrally,yang2007mimo}. 
Unfortunately, such methods are no longer valid for ISAC systems when random signals are utilized. Define $\bm{\eta}$ as the parameters of interest (POI) of the targets, e.g.,
AOA, AOD and reflection coefficients. We assume $\bm{\eta}$ remains constant in one CPI and varies between CPIs. Any estimator for $\bm{\eta}$ aims to estimate a particular realization of $\bm{\eta}$ within one CPI. Note that $\bm{\eta}$ captures  all uncertainty of $\mathbf{H}$, and when $\mathbf{H}$ is an injective map of $\bm{\eta}$, the MI between $\bm{\eta}$ and the received signals $\mathbf{y}$ is equivalent to that between $\mathbf{h}$ and $\mathbf{y}$, i.e., $I\left(\bm{\eta};\mathbf{y}|\mathbf{S}\right)=I\left(\mathbf{h};\mathbf{y}|\mathbf{S}\right)$ \cite{liu2023deterministic}. 

%In this section, we find a closed-form expression for SMI in the case that $\mathbf{h}$ is assumed to follow a Gaussian distribution, i.e., $\mathbf{h} \sim \mathcal{CN}(\mathbf{0},\mathbf{R})$. Such an assumption has been widely utilized for MIMO channel matrix \cite{1300860} and target response matrix \cite{tang2018spectrally,yang2007mimo,liu2023deterministic}. 
For ease of evaluation, we assume that $\mathbf{h}$ follows a Gaussian distribution, i.e., $\mathbf{h} \sim \mathcal{CN}(\mathbf{0},\mathbf{R})$, which has been widely adopted for MIMO channel \cite{1300860} and target response matrix \cite{tang2018spectrally,yang2007mimo,liu2023deterministic}. 
Under such circumstances, the SMI with random signals is defined as
\begin{equation}\label{SEMIdef}
	\begin{split}
		%&I_G\left(\bm{\eta};\mathbf{y}|\mathbf{S}\right)\\
		&I_G\left(\bm{\eta};\mathbf{y}|\mathbf{S}\right)=I_G\left(\mathbf{h};\mathbf{y}|\mathbf{S}\right)\\
		&\triangleq \mathbb{E}_{\mathbf{S}}\log \left\vert\mathbf{I}+ \sigma_N^{-2}\left(\mathbf{R}_R \otimes \mathbf{R}_T^{\frac{1}{2}}\mathbf{F}\mathbf{S}\mathbf{S}^\HH\mathbf{F}^\HH\mathbf{R}_T^{\frac{1}{2}}\right) \right\vert,
	\end{split}
\end{equation}
where the expectation is taken over the random signals $\mathbf{S}$. Note that $I_G\left(\bm{\eta};\mathbf{y}|\mathbf{S}\right)$ represents the highest SMI that can be achieved. Additionally, according to \cite{liu2023deterministic}, SMI is directly related to the distortion metric (e.g., the estimation error) of $\bm{\eta}$. 
	%$I_G\left(\bm{\eta};\mathbf{y}|\mathbf{S}\right) = \max_{p(\mathbf{h})}I\left(\bm{\eta};\mathbf{y}|\mathbf{S}\right)$, where $p(\mathbf{h})$ denotes the probability density function of $\mathbf{h}$. 
	%%While the communication MI between the observation and the transmit data represents the maximum achievable transmission rate, SMI has its own operational meaning. Particularly, 
	%Define $D(R)$ as the distortion-rate function for $\bm{\eta}$, which is monotonically decreasing with $R$. Then, according to \cite{liu2023deterministic}, $D(I\left(\bm{\eta};\mathbf{y}|\mathbf{S}\right))$ serves as a lower bound of the average distortion of $\bm{\eta}$. It indicates that $D(I_G\left(\bm{\eta};\mathbf{y}|\mathbf{S}\right))$ represents the minimum average distortion of $\bm{\eta}$.
	%Remarkably, the SMI is monotonically decreasing with the minimum distortion of $\bm{\eta}$ \cite{liu2023deterministic}. 
	%SMI serves as a universal lower bound for sensing	distortion metrics \cite{liu2023deterministic}.
	%where 
	%\begin{equation}\label{SEMIdefcondS}
	%	\begin{split}
		%	I_G\left(\mathbf{h};\mathbf{y}|\mathbf{S}\right)=\log \det\left(\mathbf{I}+ \sigma_N^{-2}\mathring{\mathbf{X}}^\HH\mathbf{R} \mathring{\mathbf{X}} \right),
		%	\end{split}
	%\end{equation}
	Unfortunately, the expectation in \eqref{SEMIdef} is computationally prohibitive due to the high-dimensional integrals.

\subsection{Upper bound of SMI}
An upper bound of SMI can be obtained by applying the Jensen's inequality  on \eqref{SEMIdef}. In particular, we have 
%replacing the sample covariance matrix of $\mathbf{S}\mathbf{S}^\HH$ as the statistical covariance matrix $\mathbb{E}(\mathbf{S}\mathbf{S}^\HH)=\mathbf{I}$, i.e., 
\begin{equation}
	\begin{split}
	I_G\left(\bm{\eta};\mathbf{y}|\mathbf{S}\right)\leq\log \left\vert\mathbf{I}+ \sigma_N^{-2}\left(\mathbf{R}_R \otimes \mathbf{R}_T^{\frac{1}{2}}\mathbf{F}\mathbb{E}_{\mathbf{S}}(\mathbf{S}\mathbf{S}^\HH)\mathbf{F}^\HH\mathbf{R}_T^{\frac{1}{2}}\right) \right\vert,\notag
	\end{split}
\end{equation}
where the equality holds when $\mathbf{S}\mathbf{S}^\HH = \mathbf{I}$.
Thus, we can obtain the upper bound as
\begin{equation}\label{ESEMIdefcondS_Jen}
	\begin{split}
		&\widetilde{I}_G\left(\bm{\eta};\mathbf{y}\right)
		%&\triangleq\log \det\left(\mathbf{I}+ \sigma_N^{-2}\left(\mathbf{R}_R \otimes (\mathbf{R}_T^{\frac{1}{2}}\mathbf{F}\mathbb{E}_{\mathbf{S}}(\mathbf{S}\mathbf{S}^\HH)\mathbf{F}^\HH\mathbf{R}_T^{\frac{1}{2}})\right) \right)\\
		\triangleq\log \left\vert\mathbf{I}+ \sigma_N^{-2}\left(\mathbf{R}_R \otimes \mathbf{R}_T^{\frac{1}{2}}\mathbf{F}\mathbf{F}^\HH\mathbf{R}_T^{\frac{1}{2}}\right) \right\vert.\\
	\end{split}
\end{equation}
Thus, there are two ways to interpret $\widetilde{I}_G\left(\bm{\eta};\mathbf{y}\right)$. 
%$\widetilde{I}_G\left(\bm{\eta};\mathbf{y}\right)$ has two distinct interpretations.  
On the one hand, $\widetilde{I}_G\left(\bm{\eta};\mathbf{y}\right)$ can be regarded as the MI obtained by a deterministic signal with sample covariance matrix $\mathbf{I}$. 
On the other hand, as an upper bound,  $\widetilde{I}_G\left(\bm{\eta};\mathbf{y}\right)$ can be utilized as an approximation for SMI when $N_S$ is large. This is because, as $N_S \to \infty$, the sample covariance matrix $\mathbf{S}\mathbf{S}^\HH$ will tend to be deterministic, i.e., $\lim_{N_S\to \infty} \mathbf{S}\mathbf{S}^\HH \to \mathbf{I}$. Under such circumstances, $I_G\left(\bm{\eta};\mathbf{y}|\mathbf{S}\right)$ approaches its upper bound $\widetilde{I}_G\left(\bm{\eta};\mathbf{y}\right)$. 
%Hence, the upper bound $\widetilde{I}_G\left(\bm{\eta};\mathbf{y}\right)$ can be utilized as an approximation for SMI, especially when $N_S$ is large. 
Unfortunately, as we will show later, the precision of this approximation is poor when $N_S$ is small.

\subsection{Asymptotic approximation of SMI}
In this paper, we derive a closed-form expression for the SMI by analyzing the asymptotic behavior of  $I_G\left(\mathbf{h};\mathbf{y}\right)$. In particular, we provide an approximation for the SMI in the large $N_S$ regime based on the first order approximation of the MI.  The result is shown in the following proposition.

%However, the expectation in \eqref{SEMIdef} faces the challenges caused by the high-dimensional integrals, which is computationally prohibitive. Moreover, an explicit expression is useful for performance analysis and system design.
%To address the above-mentioned issue, we use derive a tractable expression for the SEMI in the following proposition by analyzing the asymptotic behavior of $I_G\left(\mathbf{h};\mathbf{y}|\mathbf{S}\right)$.

\begin{proposition}\label{TheoRD}
	As $N_S,K \to \infty$ with a finite constant ratio $c$, i.e., $N_S/K = c$, we have
	%As $N_S\to \infty$, we have
	\begin{equation}\label{SEMI0}
		\begin{split} 
			I_G\left(\bm{\eta};\mathbf{y}|\mathbf{S}\right) =  \sum_{j=1}^{K} \bar{\varrho}_{j}(\mathbf{F}) + \mathcal{O}\left(\frac{1}{N_S}\right),
		\end{split}
	\end{equation}
%	where 
%	\begin{equation}
%		\begin{split}	
%			\bar{\varrho}_{j}=&-\log \det\left(\mathbf{I}_{N_R}+\frac{\lambda_{R,j}}{1+\lambda_{R,j}\delta(\lambda_{R,j})} \mathbf{\Lambda_{T}} \right) \\
%			&+N_T   \frac{\lambda_{R,j}\delta(\lambda_{R,j})}{1+\lambda_{R,j}\delta(\lambda_{R,j})},
%		\end{split}	
%	\end{equation}
%	$\delta(\rho)$ is the solution to the following equation
%	\begin{equation}\label{fixeq1}
%		\delta(\rho) = \frac{1}{N_T} \tr \mathbf{\Lambda_{T}}\left(\mathbf{I}_{N_R}+\frac{\rho}{1+\rho\delta(\rho)} \mathbf{\Lambda_{T}}\right)^{-1},
%	\end{equation}
%	and $\mathbf{\Lambda}_{T}=\diag([\lambda_{T,1},\cdots,\lambda_{T,N_T}])$ and $\mathbf{\Lambda}_{R}=\diag([\lambda_{R,1},\cdots,\lambda_{R,N_R}])$ with $\diag(\mathbf{x})$ denoting the diagonal martix whose diagonal is $\mathbf{x}$, and $\{\lambda_{T,i}\}_{i=1}^{N_T}$ and $\{\lambda_{R,j}\}_{j=1}^{N_R}$ denote the eigenvalues of $\mathbf{F}^{\HH}\mathbf{R}_{T}\mathbf{F}$ and $\sigma_N^{-2} \mathbf{R}_{R} $, respectively.
where 
\begin{equation}\label{averrhoj0}
	\begin{split}	
		&\bar{\varrho}_{j}(\mathbf{F})=\log \left\vert\mathbf{I}_{N_T}+\frac{\lambda_{R,j}}{1+\lambda_{R,j}\delta(\lambda_{R,j})} \mathbf{T}(\mathbf{F}) \right\vert \\
		&+N_S \log (1+\lambda_{R,j}\delta(\lambda_{R,j}))-N_S   \frac{\lambda_{R,j}\delta(\lambda_{R,j})}{1+\lambda_{R,j}\delta(\lambda_{R,j})},
	\end{split}	
\end{equation}
%\begin{equation}\label{Tdef}
%	\begin{split}
%		&\mathbf{T}(\mathbf{F})\triangleq \mathbf{R}_{T}^{\frac{1}{2}}\mathbf{F}\mathbf{F}^\HH\mathbf{R}_{T}^{\frac{1}{2}},
%	\end{split}
%\end{equation}
with $\mathbf{T}(\mathbf{F})\triangleq \mathbf{R}_{T}^{\frac{1}{2}}\mathbf{F}\mathbf{F}^\HH\mathbf{R}_{T}^{\frac{1}{2}}$. $\{\lambda_{R,j}\}_{j=1}^{K}$ denotes the eigenvalues of $\sigma_N^{-2} \mathbf{R}_{R}$, and 
$\delta(\rho)$ is the solution to the following equation
\begin{equation}\label{fixeq1}
	\delta(\rho) = \frac{1}{N_S} \tr \mathbf{T}(\mathbf{F})\left(\mathbf{I}_{N_T}+\frac{\rho}{1+\rho\delta(\rho)} \mathbf{T}(\mathbf{F})\right)^{-1}.
\end{equation}
\end{proposition}

\emph{Proof:} See Appendix \ref{proofTheoRD}.  \hfill $\blacksquare$

%\begin{remark}
%	\emph{Proposition \ref{TheoRD}} indicates that the ESMI is determined by several parameters: 1) the number of targets $K$, which indicates the number of the parameters to estimate,
%	2) the correlation at the receiver $\{\lambda_{R,j}\}_{j=1}^{N_R}$, which affects the receive SNR, 3) the eigenvalue of $\mathbf{T}(\mathbf{F})$, which represents the distribution of the transmit power, and 4) the number of frames $N_S$, which affects the sensing degree of freedom (DoF) as we will show later.
%\end{remark}

%\begin{enumerate}
%	\item The number of targets, which affects the number of the parameters to estimate.
%	\item The correlation at the receiver $\{\lambda_{R,j}\}_{j=1}^{N_R}$, which affects the receive SNR.
%	\item The eigenvalue of $\mathbf{T}(\mathbf{F})$, which represents the distribution of the transmit power.
%\end{enumerate} 

%\subsection{Sensing degree of freedom}

%The ESMI provides a novel perspective and tool for analyzing 
\emph{Proposition \ref{TheoRD}} provides a closed-form expression for the SMI between the POI and the received signals. Note that, different from the upper bound in \eqref{ESEMIdefcondS_Jen}, \emph{Proposition \ref{TheoRD}} aims to approximate SMI by analyzing its asymptotic behavior. As will be shown later, although \eqref{SEMI0} is derived under the condition that $N_S$ and $K$ approach infinity, it provides accurate approximation for SMI even when $N_S$ and $K$ are small. 
%Additionally, \eqref{SEMI0} can be a more accurate approximation for the exact value of ESMI than \eqref{ESEMIdefcondS_Jen}.
%\eqref{SEMI0} aims to calculate the exact value of ESMI by analyzing its asymptotic behavior, while \eqref{ESEMIdefcondS_Jen} measures the upper bound of ESMI.
%As we will show later, \eqref{SEMI0} can accurately approximate the ESMI even when $N_S,K$ is small, though \eqref{SEMI0} is derived under the condition that  $N_S,K \to \infty$. 
%Compared with \eqref{ESEMIdefcondS_Jen}, the approximation in \eqref{SEMI0} can be more accurate to present the exact value of ESMI, especially when $N_S$ is small.

\subsection{Effect of the number of frames}
With the explicit expression for the SMI, we can evaluate the impact of key system parameters. 
%For example, the sensing performance using random signals will be improved as $N_S$ increases. However, this has not been theoretically investigated due to the lack of a closed-form expression. 
To evaluate the effect of $N_S$, we calculate the derivative of the SMI with respect to $N_S$. 
%the previous works have not provided a theoretical analysis on the impact of $N_S$, due to the lack of an explicit expression for the sensing metric with random signals. To address this gap, we aim to investigate the effect of $N_S$ by utilizing the explicit expression of ESMI. 
%However, the theoretical analysis on the effect of $N_S$ is still unavailable in the previous works due to the lack of explicit expression for the sensing metric with random signals. To fill this blank, we will study the effect of $N_S$ based on the explicit expression of ESMI. 
First, for all $N_S\geq 0$, the derivative of $\bar{\varrho}_{j}(\mathbf{F})$, which is given in \eqref{averrhoj0}, with respect to $N_S$ can be given by 
\begin{equation}
	\begin{split}
		&\frac{\partial \bar{\varrho}_{j}}{\partial N_S}=-\frac{\lambda_{R,j}^2 \tr\mathbf{T}\left(\mathbf{I}_{N_T}+\frac{\lambda_{R,j}}{1+\lambda_{R,j}\delta(\lambda_{R,j})} \mathbf{T} \right)^{-1}}{(1+\lambda_{R,j}\delta(\lambda_{R,j}))^2} \\
		&\quad  +\frac{N_S\lambda_{R,j}}{1+\lambda_{R,j}\delta(\lambda_{R,j})}-\frac{N_S\lambda_{R,j}}{(1+\lambda_{R,j}\delta(\lambda_{R,j}))^2}\\
		&\quad 
		+\log (1+\lambda_{R,j}\delta(\lambda_{R,j}))- \frac{\lambda_{R,j}\delta(\lambda_{R,j})}{1+\lambda_{R,j}\delta(\lambda_{R,j})} \\
		&\overset{(a)}{=}\log (1+\lambda_{R,j}\delta(\lambda_{R,j}))- \frac{\lambda_{R,j}\delta(\lambda_{R,j})}{1+\lambda_{R,j}\delta(\lambda_{R,j})} \geq 0,
	\end{split}
\end{equation}
where step (a) follows \eqref{fixeq1}. Thus, we have \begin{equation}\label{gradientNSall}
\frac{\partial I_G\left(\bm{\eta};\mathbf{y}|\mathbf{S}\right)}{\partial N_S}= \sum_{j=1}^K\frac{\partial \bar{\varrho}_{j}}{\partial N_S}\geq 0.
\end{equation}
This indicates that the SMI $I_G\left(\bm{\eta};\mathbf{y}|\mathbf{S}\right)$ is monotonically increasing with respect to $N_S$.

\subsection{Sensing DoF loss}
%To further quantify the effect of $N_S$, 
To measure the impact of random signals on sensing, we further investigate the sensing Degree of Freedom (DoF). This concept was introduced in \cite{xiong2023fundamental} to measure the loss of CRB induced by the random signals and defined as the effective number of independent observations.
%Intuitively, the sensing DoF should be defined as the number of ``effective'' independent observations \cite{xiong2023fundamental}. 
By following the same idea, we define the SMI-oriented sensing DoF as
\begin{equation}\label{sdofdef}
	\nu_s =  \lim_{\sigma_N^2\to 0}  \frac{N_S I_G\left(\bm{\eta};\mathbf{y}|\mathbf{S}\right)}{\widetilde{I}_G\left(\bm{\eta};\mathbf{y}|\mathbf{S}\right)}.
\end{equation}
where the maximum sensing DoF is the total number of the independent observations $N_S$.
The normalization coefficient $\frac{I_G\left(\mathbf{h};\mathbf{y}\right)}{\widetilde{I}_G\left(\mathbf{h};\mathbf{y}\right)}$ is the ratio between the SMI with random signals and its upper bound achieved by deterministic signals. Thus, this ratio measures the MI loss caused by the randomness of signals. The following proposition provides the lower bound for $I_G\left(\bm{\eta};\mathbf{y}|\mathbf{S}\right)$, based on which the maximum sensing DoF loss can be obtained. 
%To find the maximum of the sensing DoF loss, we provide a lower bound for $I_G\left(\mathbf{h};\mathbf{y}\right)$ in the following proposition.
\begin{proposition}\label{TheoIlb}
	When $N_S \geq K$, we have
	\begin{equation}\label{Ilower bound}
		\begin{split} \! \!  I_G\left(\bm{\eta};\mathbf{y}|\mathbf{S}\right) \geq \frac{N_S - \min\{K,\mathrm{rank}(\mathbf{F}\mathbf{F}^\HH)\}}{N_S} \widetilde{I}_G\left(\bm{\eta};\mathbf{y}|\mathbf{S}\right).
		\end{split}
	\end{equation}
\end{proposition}

\emph{Proof:} See Appendix \ref{proofTheoIlb}. \hfill $\blacksquare$

Based on \emph{Proposition \ref{TheoIlb}}, the sensing DoF can be bounded by 
\begin{equation}\label{SDoF_range}
	N_S - \min\{K,\mathrm{rank}(\mathbf{F}\mathbf{F}^\HH)\} \leq \nu_s \leq N_S.
\end{equation}
This indicates that $\min\{K,\mathrm{rank}(\mathbf{F}\mathbf{F}^\HH)\}$ is the maximum sensing DoF loss defined based on SMI, which coincides with the sensing DoF loss defined based on CRB \cite{xiong2023fundamental}. 
%As the value of $K$ increases, the sensing task becomes more onerous due to various factors, e.g., the presence of the interference and the need for higher resolution. This necessitates a large number of samples to effectively estimate POI. 
%In particular, more samples enables the system to obtain a more accurate estimation of POI. 
%In particular, more samples provide a higher level of redundancy, enabling the system to average out the random variations and obtain a more reliable estimate of the target's characteristics. 
It can be observed that maximum sensing DoF loss will increase with $K$. This is because the system needs to separate the signals associated with different targets, and this task becomes more challenging as the number of targets increases, due to the interference between different targets. 
%As the value of $K$ increases, the sensing task becomes more onerous. 
%In particular, the system needs to separate the signals associated with each target from the overall received signals. This task becomes more challenging as the number of targets increases, due to the interference between different targets. Therefore, the maximum sensing DoF loss will increase. 
As a result, more samples are required to mitigate such a loss. 
In particular, according to the squeeze theorem, there is no loss when $N_S \to \infty$, because the lower bound of $I_G\left(\bm{\eta};\mathbf{y}|\mathbf{S}\right)$ in \eqref{Ilower bound} approaches $\widetilde{I}_G\left(\bm{\eta};\mathbf{y}|\mathbf{S}\right)$. This agrees with the theoretical results in \eqref{gradientNSall}.

\section{SMI-Oriented Precoding Design}

Next, we will optimize the precoder to maximize the SMI. The optimization problem is formulated as
\begin{equation}\label{P0}
	\begin{array}{ccl}
		\mathcal{P}_0:&\max\limits_{\mathbf{F}} &\mathcal{L}\left(\mathbf{F}\right)\\
		& s.t. & \Vert\mathbf{F}\Vert^2 \leq P,
	\end{array}
\end{equation}
where $\mathcal{L}\left(\mathbf{F}\right)=\sum_{j=1}^{K} \bar{\varrho}_{j}(\mathbf{F})$ and  the transmit power is constrained to a maximum value of $P$. 
A common practice to solve $\mathcal{P}_0$ is to utilize the interior point method associated with the Newton's method, which requires both gradient and Hessian matrix. However, since $\delta$ is obtained by solving the equation defined in \eqref{fixeq1},  it is difficult to obtain the Hessian matrix of $\mathcal{L}(\mathbf{F})$ with respect to $\mathbf{F}$. To address this issue, we propose to solve the problem $\mathcal{P}_0$ by exploiting the manifold steepest descent method, which makes it easier to deal with the constraint and only requires the gradient. 
%Note that $\delta$ is obtained by solving the equation defined in \eqref{fixeq1}, which is dependent on $\mathbf{F}$. This makes the optimization problems in \eqref{P0} hard to solve, especially with the non-convex constraint. To address this issue, we propose to solve the problems $\mathcal{P}_0$ by exploiting the manifold steepest descent method, which makes the constraint easier to deal with.

The manifold-based method updates the variable within the
tangent space. By updating along the tangent space with
a small enough step, the new point is almost within the manifold.
The manifold steepest descent method requires the Euclidean gradient, the Riemannian gradient, and the projections on the manifold.

\subsubsection{Euclidean Gradient}
Define $\nabla_{\mathbf{F}}	\mathcal{L}\left(\mathbf{F}\right)$ as the Euclidean gradient of the objective function in \eqref{P0} with respect to $\mathbf{F}$, which is computed as $\nabla_{\mathbf{F}}	\mathcal{L}\left(\mathbf{F}\right) =\sum_{j=1}^{K}  \nabla_{\mathbf{F}}\bar{\varrho}_{j}(\mathbf{F})$, 
%\begin{equation}
%	\begin{split}
%		\nabla_{\mathbf{F}}	\mathcal{L}\left(\mathbf{F}\right) =\sum_{j=1}^{K}  \nabla_{\mathbf{F}}\bar{\varrho}_{j}(\mathbf{F}),
%	\end{split}
%\end{equation}
with $\nabla_{\mathbf{F}}\bar{\varrho}_{j}(\mathbf{F})$ denoting the Euclidean gradient of $\bar{\varrho}_{j}(\mathbf{F})$ with respect to $\mathbf{F}$. Note that $\delta$ defined in \eqref{fixeq1} is dependent on $\mathbf{F}$, which makes the derivative more complex. To this end, we first give the following lemma for obtaining the derivatives of $\delta(\rho)$ with respect to $\mathbf{F}$.
%, which is defined by
%\begin{equation}
%	\mathbf{\Delta}_{\rho}'(\mathbf{F})=\frac{\partial \delta(\rho)}{\partial \mathbf{F}^*}.
%\end{equation}

\begin{proposition}\label{lemmaderiveRs}
	Define $\alpha(\rho)=\frac{\rho}{1+\rho\delta(\rho)}$. Then, the gradient of $\delta(\rho)$ with respect to $\mathbf{F}$, i.e., $\mathbf{\Delta}_{\rho}'\triangleq
	\frac{\partial \delta(\rho)}{\partial \mathbf{F}^*}$ is obtained by
	\begin{equation}\label{derivaLER}
	\begin{split}
		&\mathbf{\Delta}_{\rho}'  = \frac{\mathbf{R}_T^{\frac{1}{2}}\left(\mathbf{I}+\alpha(\rho) \mathbf{T}(\mathbf{F})\right)^{-2}\mathbf{R}_T^{\frac{1}{2}}\mathbf{F}}{N_S-\alpha^2(\rho)\tr\left(\mathbf{T}(\mathbf{F}) \left(\mathbf{I}+\alpha(\rho) \mathbf{T}(\mathbf{F})\right)^{-1}\right)^2 }.
	\end{split}
\end{equation}
	%The $(m,n)$th entry of $\mathbf{\Delta}_{\rho}'(\mathbf{F})$, denoted as $[\mathbf{\Delta}_{\rho}'(\mathbf{F})]_{m,n}$, can be obtained by 
%	\begin{equation}\label{derivaLER}
%		\begin{split}
%				[\mathbf{\Delta}_{\rho}'(\mathbf{F})]_{m,n} = \frac{[\mathbf{R}_T^{\frac{1}{2}}\mathbf{M}_T^2(\rho)\mathbf{R}_T^{\frac{1}{2}}\mathbf{F}]_{m,n}}{N_S-\alpha^2(\rho)\tr\left(\mathbf{T}(\mathbf{F}) \mathbf{M}_{T}(\rho)\right)^2 }
%		\end{split}
%	\end{equation}
%	where $\mathbf{M}_{T}(\rho) =\left(\mathbf{I}+\frac{\rho}{1+\rho\delta(\rho)} \mathbf{T}(\mathbf{F})\right)^{-1}$.
%	\begin{equation}\label{derivaLER}
%	\begin{split}
%		&[\mathbf{\Delta}_{\rho}'(\mathbf{F})]_{m,n} \\
%		&= \frac{\left[\mathbf{R}_T^{\frac{1}{2}}\left(\mathbf{I}+\frac{\rho}{1+\rho\delta(\rho)} \mathbf{T}(\mathbf{F})\right)^{-2}\mathbf{R}_T^{\frac{1}{2}}\mathbf{F}\right]_{m,n}}{N_S-\alpha^2(\rho)\tr\left(\mathbf{T}(\mathbf{F}) \left(\mathbf{I}+\frac{\rho}{1+\rho\delta(\rho)} \mathbf{T}(\mathbf{F})\right)^{-1}\right)^2 }.
%	\end{split}
%\end{equation}
\end{proposition}

\emph{Proof:} See Appendix \ref{prooflemmaderiveRs}. \hfill $\blacksquare$

Based on \emph{Proposition \ref{lemmaderiveRs}},  $\nabla_{\mathbf{F}}\bar{\varrho}_{j}(\mathbf{F})$ can be obtained by\footnote{Note that we simplify the notation by omitting the argument of $\mathbf{M}_{T,j}(\mathbf{F})$, $\alpha_j(\mathbf{F})$, $\mathbf{T}(\mathbf{F})$, and $\mathbf{\Delta}_{\lambda_{R,j}}'(\mathbf{F})$ in \eqref{varrhograd}.}
%	\begin{equation}\label{varrhograd}
%		\begin{split}
%			&\nabla_{\mathbf{F}}\bar{\varrho}_{j}(\mathbf{F})%=\frac{\partial \bar{\varrho}_{j}(\mathbf{F})}{\partial \mathbf{F}^{*}}\\
%			%&
%			=\alpha_j^2\tr\left(\mathbf{M}_{T,j}\mathbf{T}\right)\mathbf{\Delta}_{\lambda_{R,j}}'+ \alpha_j\tr\left(\mathbf{M}_{T,j}\right)\mathbf{R}_T \mathbf{F} \\
%			&+N_S\alpha_j \mathbf{\Delta}_{\lambda_{R,j}}'-\frac{N_S}{\left(1+\lambda_{R,j}\delta(\lambda_{R,j})\right)^2} \mathbf{\Delta}_{\lambda_{R,j}}'
%		\end{split}
%	\end{equation}
\begin{align}
	&\nabla_{\mathbf{F}}\bar{\varrho}_{j}(\mathbf{F})=\alpha_j^2\tr\left(\mathbf{M}_{T,j}\mathbf{T}\right)\mathbf{\Delta}_{\lambda_{R,j}}'+ \alpha_j\tr\left(\mathbf{M}_{T,j}\right)\mathbf{R}_T \mathbf{F}\notag \\
	&+N_S\alpha_j \mathbf{\Delta}_{\lambda_{R,j}}'-\frac{N_S}{\left(1+\lambda_{R,j}\delta(\lambda_{R,j})\right)^2} \mathbf{\Delta}_{\lambda_{R,j}}', \label{varrhograd}
\end{align}
	where $\alpha_j(\mathbf{F}) = \alpha(\lambda_{R,j})$ and 
%	$\mathbf{M}_{T,j}(\mathbf{F})=\left(\mathbf{I}_{N_S}+\alpha_j(\mathbf{F}) \mathbf{R}_{T}^{\frac{1}{2}}\mathbf{F}\mathbf{F}^\HH\mathbf{R}_{T}^{\frac{1}{2}} \right)^{-1}.$
	\begin{equation}
		\begin{split}
			\mathbf{M}_{T,j}(\mathbf{F})=\left(\mathbf{I}_{N_S}+\alpha_j(\mathbf{F}) \mathbf{R}_{T}^{\frac{1}{2}}\mathbf{F}\mathbf{F}^\HH\mathbf{R}_{T}^{\frac{1}{2}} \right)^{-1}.
		\end{split}
	\end{equation}

\subsubsection{Riemannian Gradient on Manifold}
%We first reformulate the constraint (\ref{P0}) as the manifold.
%We define the manifold $\mathcal{M}_{\mathbf{F}}$ as
%\begin{equation}
%	\mathcal{M}_{\mathbf{F}}=\left\{\mathbf{F}\in \mathcal{C}^{K\times N_S}\Big|\Vert\mathbf{F}\Vert^2 = P \right\}.
%\end{equation}
%The manifold $\mathcal{M}_{\mathbf{F}}$ is also known as the complex oblique manifold. 
%The tangent space at a given point $\mathbf{F}\in\mathcal{M}_{\mathbf{F}}$  is given by
%\begin{equation}
%	\mathcal{T}_{\mathbf{U}} \mathcal{M}_{\mathbf{F}} =\left\{\mathbf{U}\in \mathcal{C}^{K\times N_S}\Big|  \tr\left(\mathbf{F}\mathbf{U}^\HH+\mathbf{U}\mathbf{F}^\HH\right)=\mathbf{0}\right\}.
%\end{equation}
%The Riemannian gradient at one point $\mathbf{F}\in\mathcal{M}_{\mathbf{F}}$ can be seen as the projection of the Euclidean gradient onto the tangent space $T_{\mathbf{U}} \mathcal{M}_{\mathbf{F}}$ at that point.
%Therefore, the Riemannian gradient of
For the manifold $\mathcal{M}_{\mathbf{F}}=\left\{\mathbf{F}\in \mathcal{C}^{K\times N_S}\Big|\Vert\mathbf{F}\Vert^2 = P \right\}$, the Riemannian gradient of $\mathcal{L}\left(\mathbf{F}|\mathbf{v}_{k},\mathbf{Z}_{k}\right)$ at $\mathbf{F}\in\mathcal{M}_{\mathbf{F}}$ is given by \cite{xie2022perceptive}
\begin{equation}\label{RiemGradF}
	\begin{split}
		&\mathrm{grad}_{\mathbf{F}} \mathcal{L}\left(\mathbf{F}\right)=\nabla_{\mathbf{F}} \mathcal{L}\left(\mathbf{F}\right)-\left(\frac{1}{P}\tr\Re\left(\mathbf{F}\nabla_{\mathbf{F}}^\HH \mathcal{L}\left(\mathbf{F}\right)\right)\right)\mathbf{F}.
	\end{split}
\end{equation}

\subsubsection{Retraction}
A retraction is needed to remap the updated points from the tangent space onto the manifold. The retraction of a tangent matrix $\mathbf{D}\in T_{\mathbf{U}} \mathcal{M}_{\mathbf{F}}$ at $\mathbf{F}$ is defined as 
\begin{equation}\label{retractionF}
	\begin{split}
%		\mathcal{P}_{\mathbf{F}}:&\mathcal{T}_{\mathbf{U}} \mathcal{M}_{\mathbf{F}} \to  \mathcal{M}_{\mathbf{F}}\\
%		& \mathbf{D} \to
%		\frac{P}{\Vert\mathbf{F}+\mathbf{D}\Vert}\left(\mathbf{F}+\mathbf{D}\right).
	\mathcal{P}_{\mathbf{F}}(\mathbf{D})=
\frac{P}{\Vert\mathbf{F}+\mathbf{D}\Vert}\left(\mathbf{F}+\mathbf{D}\right).
	\end{split}
\end{equation}

To solve (\ref{P0}), we propose the manifold-based method summarized in \textbf{Algorithm 1}, whose convergence is guaranteed by \cite[Theorem 4.3.1]{absil2009optimization}.

\begin{algorithm}[h] 
	\caption{Proposed Manifold-based Method to optimize $\mathbf{F}^{\dagger}$} 
	\textbf{Input:} An initial point $\mathbf{F}_{0}$, $\mathbf{D}_{0}=-\mathrm{grad}_{\mathbf{F}} \mathcal{L}\left(\mathbf{F}_{0}\right)$, and $m=0$.
	
	\textbf{Repeat}
	\begin{enumerate} 
		\item Compute $\beta_{m}$ via the Armijo line search step \cite[Definition 4.2.2]{absil2009optimization}.
		\item Update $\mathbf{F}_{m+1}=\mathcal{P}_{\mathbf{F}}(-\beta_{m} \mathrm{grad}_{\mathbf{F}} \mathcal{L}\left(\mathbf{F}_{m}\right))$ via (\ref{RiemGradF}) and (\ref{retractionF}).
		\item $m \gets m+1$.
	\end{enumerate} 
	\textbf{Until} Convergence criterion is met.\\
	\textbf{Output:} The optimal solution $\mathbf{F}^{\dagger}=\mathbf{F}_{m}$.
	\label{ALG1}
\end{algorithm}

\section{Simulation Result}
In this section, we will validate the accuracy of the theoretical analysis and the effectiveness of proposed manifold-based algorithms by simulations. We consider a mmWave system operating at a carrier frequency of 28 GHz. The AOD and AOA of the targets are generated uniformly in the range  $[30^\circ,60^\circ]$. The number of antennas on the transmitter and receiver is set as $N_T = 32$ and $N_R =16$, respectively. 
%The number of frames in one pulse is $N_s=128$, unless otherwise specified.
%The noise power is $\sigma^2=-90$ dBm. 
For the manifold optimization, we set the maximum number of iterations as $50$. To terminate the iteration, the tolerance for the norm of the gradient between two iterations is $10^{-5}$. The transmission power is set as $P = 30$ dBm, the noise power is $\sigma^2=-90$ dBm, and the signal-to-noise ratios (SNR) is set as $\mathrm{SNR} \triangleq \frac{P\sigma_{k}^2}{\sigma^2}=20$ dB. 
%the variance of $\epsilon_{k}$ is set as $\sigma_{k}^2 = 10^{-10}$. %The signal-to-noise ratios (SNR) is set as $20$ dB.

\subsection{Validation of the SMI}

\begin{figure}[!t]
	\centering
	\includegraphics[width=2.8in]{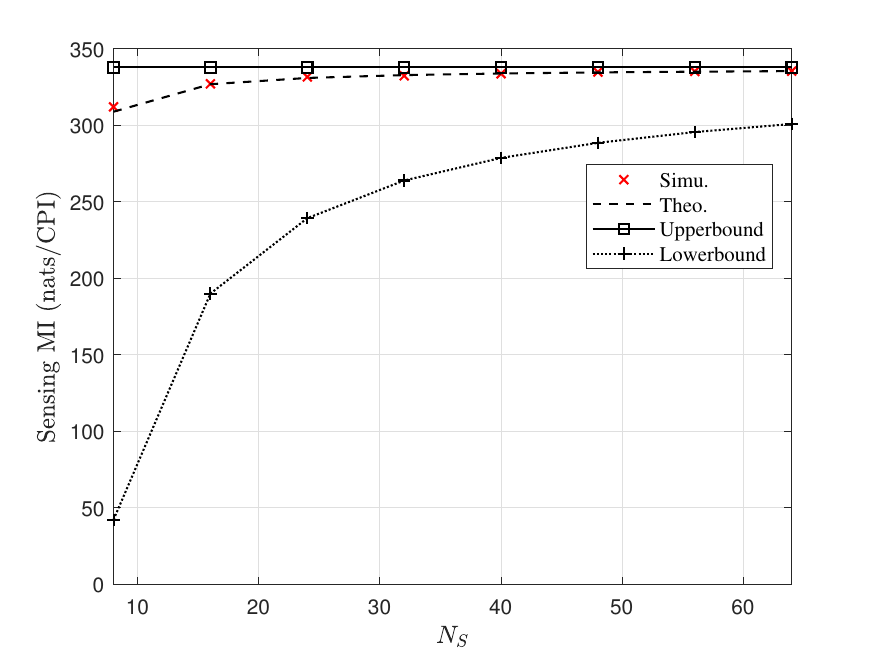}
	\caption{Sensing MI versus the number of frames $N_S$. %(Averaged over 5000 Monte-Carlo trials)
	}
	\label{fig_fitness}
\end{figure}

Fig. \ref{fig_fitness} validates the accuracy of the evaluation for SMI in \emph{Proposition \ref{TheoRD}} with $K=7$ targets. The legend `Simu.' denotes the MI obtained by Monte-Carlo simulations, i.e., 
$\bar{I}\left(\bm{\eta};\mathbf{y}|\mathbf{S}\right)=\frac{1}{N_{mc}}\sum_{i=1}^{N_{mc}}I_G\left(\bm{\eta};\mathbf{y}|\mathbf{S}=\mathbf{S}_{i}\right),$ 
%\begin{equation}
%	\begin{split}
%		\bar{I}\left(\bm{\eta};\mathbf{y}|\mathbf{S}\right)=\frac{1}{N_{mc}}\sum_{i=1}^{N_{mc}}I_G\left(\bm{\eta};\mathbf{y}|\mathbf{S}=\mathbf{S}_{i}\right),
%	\end{split}
%\end{equation}
where %$I_G\left(\mathbf{h};\mathbf{y}|\mathbf{S}\right)$ is defined by \eqref{SEMIdefcondS}, 
$\mathbf{S}_{i}$ denotes the $i$th realization of $\mathbf{S}$ and $N_{mc} = 5000$ represents the number of Monte-Carlo trails. 
%The number of targets is set as $K=7$. 
%The number of targets is $K=2$ and the number of receive antennas is $N_R = 4$. Here, we set $N_T$ and $N_S$ increase together with a fixed ratio, i.e., $N_T = 2N_S$. 
%The number of frames is set as $N_S=8$.
The legend `Theo.' denotes the theoretical result of the SMI given in \eqref{SEMI0}. 
The legend `Upperbound' denotes the upper bound in \eqref{ESEMIdefcondS_Jen} and the legend `Lowerbound' represents the lower bound of SMI given in \emph{Proposition \ref{TheoIlb}}. 
It can be observed that the approximation in \eqref{ESEMIdefcondS_Jen} is very accurate, particularly when $N_S$ is large. This validates \emph{Proposition \ref{TheoRD}}. 
Meanwhile, a discrepancy consistently exists between SMI and its upper bound. 
As $N_S$ decreases, this discrepancy becomes more obvious.  %This observation is also our motivation to present the tractable expression for the SMI. 

%\begin{figure}[!t]
%	\centering
%	\includegraphics[width=3.5in]{SEMI_fig_0dB_N_T_test1.eps}
%	\caption{SEMI versus the number of transmit antennas $N_T$. 
%	}
%	\label{fig_fitness_N_T}
%\end{figure}

%\begin{figure}[!t]
%	\centering
%	\includegraphics[width=3.5in]{SEMI_fig_N_T_3_1.eps}
%	\caption{ESMI versus the number of transmit antennas $N_T$. 
%	}
%	\label{fig_fitness_N_T}
%\end{figure}
%
%Fig. \ref{fig_fitness_N_T} shows the comparison between ‘Simu.’ and ‘Theo.’ results versus the number of transmit antennas $N_T$. We can observe that, 
%as $N_T$ increases, the discrepancy between the exact value of the average sensing mutual information and its upper bound becomes more pronounced. This phenomenon can be attributed to the fact that the requirement of samples increases with the growth of $N_T$. This observation suggests that the J-ESMI necessitates a significantly large number of samples to accurately represent the actual average performance, potentially leading to high latency. This scenario is further exacerbated by the deployment of extremely large-scale antenna arrays, which is one prominent characteristic of several cutting-edge technologies being considered for the next generation of communication system \cite{9903389}.

\begin{figure}[!t]
	\centering
	\includegraphics[width=2.8in]{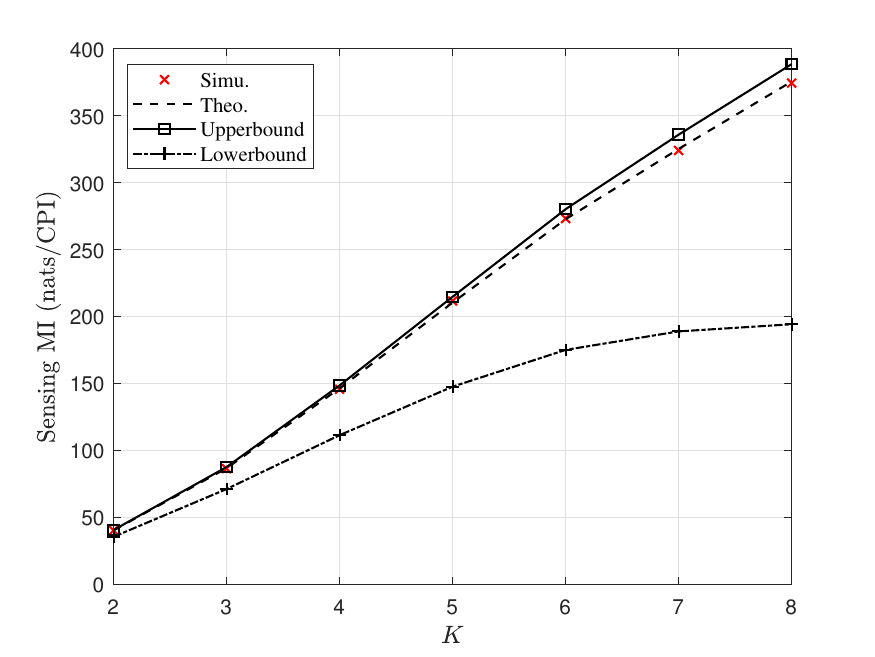}
	\caption{Sensing MI versus the number of targets $K$. 
	}
	\label{fig_fitness_K}
\end{figure}

Fig. \ref{fig_fitness_K} compares the theoretical and simulation results with $N_S=32$. 
We can observe that, as $K$ increases, the discrepancy between the SMI and its upper bound becomes more pronounced. This is because the sensing DoF loss increases with the increasing of $K$. In this case, more samples are required to approach the upper bound of SMI, potentially leading to high latency. 

\subsection{Performance of the proposed manifold-based method}

%\begin{figure}[!t]
%	\centering
%	\includegraphics[width=3.5in]{SEMI_iter_8.eps}
%	\caption{Ergodic sensing MI over iterations.}
%	\label{fig_iter}
%\end{figure}

%We begin with evaluating the convergence behavior of the proposed manifold-based algorithm. The number of frames is set as $N_S=8$. The legend ‘Proposed Method’ denotes the ESMI obtained by the proposed manifold-based method.
%The legend ‘Water-Filling’ denotes the ESMI obtained by the method given in \cite{biguesh2006training}, which aimed at optimizing the J-ESMI. Fig. \ref{fig_iter} plots the objective value of the problem (36). From Fig. \ref{fig_iter}, we can see that the manifold-based algorithm can converge after several iterations. Moreover, we can see that the proposed ESMI-oriented optimization can outperform the water-filling optimization based on the J-ESMI. This is because the J-ESMI fails to evaluate the exact value of ESMI when $N_S$ is not large enough. Optimizing J-ESMI can not effectively improve the actual ESMI.

\begin{figure}[!t]
	\centering
	\includegraphics[width=2.8in]{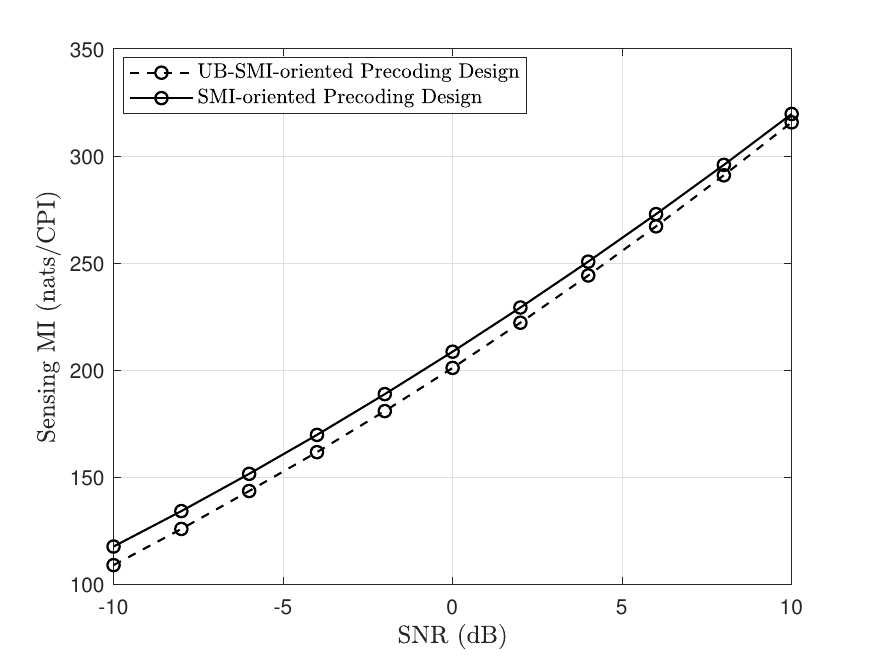}
	\caption{Sensing MI versus SNR.}
	\label{fig_comp}
\end{figure}

%We compare the proposed method with the water-filling method given in \cite{biguesh2006training}, which aimed at optimizing the DSMI. We set $N_S =32$. 
Note that, as $N_S \to \infty$, SMI will approach its upper bound \eqref{ESEMIdefcondS_Jen}. This allows us to utilize the upper bound as an approximation when $N_S$ is large. However, when $N_S$ is small, the approximation is not accurate.
%In this case, it can potentially result in a performance loss to optimize the upper bound for precoder design. To validate this, 
In Fig. \ref{fig_comp}, we compare the proposed method that maximizes the SMI in \eqref{SEMI0} and the method that optimizes the upper bound in \eqref{ESEMIdefcondS_Jen}. 
%In Fig. \ref{fig_comp}, 
The legend ‘SMI-oriented Precoding Design’ denotes the SMI obtained by the proposed manifold-based method and the legend ‘UB-SMI-oriented Precoding Design’ represents the SMI obtained by the method given in \cite{biguesh2006training}, which maximizes \eqref{ESEMIdefcondS_Jen}. 
The number of frames and targets are set as $N_S =32$ and $K=15$, respectively. 
%Figure \ref{fig_comp} illustrates the obtained ESMI using the water-filling method and the proposed method at varying SNR. 
It can be observed from Fig. \ref{fig_comp} that the proposed method outperforms the existing one, and the performance gap between the two methods becomes smaller as SNR increases. %This can be attributed to the fact that the mismatch between the DSMI and the actual ergodic sensing mutual information diminishes with the increasing of SNR. 

\section{Conclusion}
This paper investigated the achievable sensing performance and precoder design for ISAC systems utilizing random signals. For that purpose, we first derived the closed-form expression for SMI by utilizing random matrix theory. 
It was shown that the SMI with random signals is upper bounded by that with specific deterministic signals and SMI approaches its upper bound when the number of frames increases to infinity.  Additionally, we analyzed the effect of the number of sensing frames and investigated the sensing DoF loss caused by random signals.  
%Additionally, it was shown that the SMI with random signals is upper bounded by the SMI with specific deterministic signals, and SMI approaches its upper bound when the number of frames increases to infinity. 
Finally, a manifold-based optimization approach was proposed to maximize the SMI by designing the precoder. 
Simulation results validated the accuracy of the theoretical result and the effectiveness of the proposed precoder design method.  
	
	\appendices
	\section{Proof of Proposition \ref{TheoRD}}\label{proofTheoRD}
	This proof is equivalent to analyze the asymptotic behavior of the following random variable
	\begin{equation}\label{SEMI00}
		\begin{split} 
			\varrho =\log \left\vert\mathbf{I}+ \sigma_N^{-2}\mathbf{R}_R \otimes (\mathbf{S}^\HH\mathbf{F}^\HH\mathbf{R}_T \mathbf{F}\mathbf{S}) \right\vert.
	\end{split}
	\end{equation}
In fact, \emph{Proposition} \ref{TheoRD} is an extension of 
 \cite[Theorems 1]{hachem2008new}, which is summarized in the following lemma. 
 
 	\begin{lemma}[{\cite[Theorems 1]{hachem2008new}}]
  Given two diagonal matrices $\mathbf{D}_n$ and $\widetilde{\mathbf{D}}_n$, we can define $\mathbf{Z}_n=\mathbf{D}_n^{\frac{1}{2}} \widetilde{\mathbf{Z}}_n \widetilde{\mathbf{D}}_n^{\frac{1}{2}}$, where $\widetilde{\mathbf{Z}}_n$ has i.i.d. entries with distribution $\mathcal{CN}(0,\frac{1}{n})$. Then, we have 
 \begin{equation}\label{oldlemma}
 	\begin{split}
 		\mathbb{E}  \log \left\vert\rho \mathbf{Z}_n\mathbf{Z}_n^\HH+\mathbf{I}\right\vert = V_{n}(\rho) + \mathcal{O}\left(\frac{1}{n}\right),
 	\end{split}
 \end{equation}
 where  
 \begin{equation}\label{Vnrhodef}
 	\begin{split}
 		V_{n}(\rho) = & \log \left\vert\mathbf{I}+\rho \delta_n(\rho) \widetilde{\mathbf{D}}_n\right\vert + \log \left\vert\mathbf{I}+\rho \tilde{\delta}_n(\rho) \mathbf{D}_n\right\vert \\
 		&-n \rho \delta_n(\rho)\tilde{\delta}_n(\rho),
 	\end{split}
 \end{equation}
 and  $(\delta_n(\rho),\tilde{\delta}_n(\rho))$ is the unique positive solution of the following fixed-point equations
 \begin{equation}\label{oldfp}
 	\begin{split}
 		\left\{
 		\begin{matrix}
 			\delta = \frac{1}{n} \tr  \mathbf{D}_n\left(\mathbf{I}+\rho \tilde{\delta} \mathbf{D}_n\right)^{-1}\\
 			\tilde{\delta} = \frac{1}{n} \tr \widetilde{\mathbf{D}}_n\left(\mathbf{I}+\rho \delta  \widetilde{\mathbf{D}}_n\right)^{-1}
 		\end{matrix}\right. .
 	\end{split}
 \end{equation}	
\end{lemma}

 The main idea of this proof is to recast $I_G\left(\mathbf{h};\mathbf{y}|\mathbf{S}\right)$ into an extended model which fits into the framework of \cite{hachem2008new}.

First, by utilizing the property of block diagonal matrix and the equality  $\det(\mathbf{I}+\mathbf{A}\mathbf{B})=\det(\mathbf{I}+\mathbf{B}\mathbf{A})$, (\ref{SEMI00}) can be rewritten as 
$\varrho =\sum_{j=1}^{K} \varrho_j$, 
%\begin{equation}\label{rhojsum}
%	\varrho =\sum_{j=1}^{K} \varrho_j,
%\end{equation}
where $\varrho_j=\log \det\left(\mathbf{I}_{N_T} + \lambda_{R,j} \mathbf{R}_{T}^{\frac{1}{2}}\mathbf{F}\mathbf{S}\mathbf{S}^\HH \mathbf{F}^\HH\mathbf{R}_{T}^{\frac{1}{2}} \right).$ 
%By performing the eigen-value decomposition on $\mathbf{D}$, and $\mathbf{T}$, we have $\mathbf{D}=\mathbf{U}_D \widehat{\mathbf{D}} \mathbf{U}_D^\HH$ and $\mathbf{T}=\mathbf{U}_T \widehat{\mathbf{T}} \mathbf{U}_T^\HH$, respectively. Note that $\widehat{\mathbf{D}}$ and $\widehat{\mathbf{T}}$ are diagonal and $\mathbf{U}_D$ and $\mathbf{U}_T$ are unitary. 
By performing the eigenvalue decomposition on $\mathbf{T}\triangleq \mathbf{R}_{T}^{\frac{1}{2}}\mathbf{F}\mathbf{F}^\HH\mathbf{R}_{T}^{\frac{1}{2}}$, we have $\mathbf{T}=\mathbf{U}_T \widetilde{\mathbf{\Lambda}}_T \mathbf{U}_T^\HH$, where $\mathbf{U}_T \in \mathcal{C}^{N_T\times N_T}$ denotes the eigenvectors and
\begin{equation}
	\begin{split}
		\widetilde{\mathbf{\Lambda}}_T=\left[\begin{matrix}
		\mathbf{\Lambda}_T & \mathbf{0}\\
		\mathbf{0} & \mathbf{0}
		\end{matrix}
	\right],
	\end{split}
\end{equation}
with $\mathbf{\Lambda}_T \in \mathcal{C}^{K\times K}$ denotes the eigenvalue matrix.

Note that, with any unitary matrices $\mathbf{U}$ and $\mathbf{V}$, the random matrices $\mathbf{U} \mathbf{S} \mathbf{V}$ and $\mathbf{S}$ are statistically equivalent \cite{hachem2008new}. Thus, by omitting some constants independent of $\mathbf{S}$, the asymptotic behavior of $\varrho_j$ is equivalent to that of the RV $\hat{\varrho}_j = \log \det\left(\mathbf{I}_{K} + \lambda_{R,j} \mathbf{\Lambda}_{T}^{\frac{1}{2}}\mathbf{S}\mathbf{S}^\HH \mathbf{\Lambda}_{T}^{\frac{1}{2}} \right)$.
By invoking $\mathbf{D}_n=\mathbf{\Lambda}_{T}$, $\widetilde{\mathbf{D}}_n=\mathbf{I}_{N_S}$, $n=N_S$, and $\rho = \lambda_{R,j}$ into \eqref{oldfp}, we have $\tilde{\delta}(\rho)=1/(1+\rho \delta )$ such that the fixed-point equations in \eqref{oldfp} reduce to a linear one with respect to $\delta$, i.e.,
\begin{equation}\label{deltadef1}
	\delta(\rho) = \frac{1}{N_S} \tr \mathbf{\Lambda_{T}}\left(\mathbf{I}_{K}+\frac{\rho}{1+\rho\delta(\rho)} \mathbf{\Lambda_{T}}\right)^{-1}.
\end{equation}
Furthermore, from \emph{Lemma 1}, we can obtain that, as $N_S,N_T \to \infty$, one has
\begin{equation}\label{exprhoj}
	\begin{split}
		\mathbb{E}(\hat{\varrho}_j) =\bar{\varrho}_j+\mathcal{O}\left(\frac{1}{N_S}\right), j =1,2,\cdots,
	\end{split}
\end{equation}
where 
\begin{equation}\label{averrhoj1}
	\begin{split}
		&\bar{\varrho}_j = \log \left\vert\mathbf{I}_{K}+\frac{\lambda_{R,j}}{1+\lambda_{R,j}\delta(\lambda_{R,j})} \mathbf{\Lambda_{T}} \right\vert\\
		&+N_S \log (1+\lambda_{R,j}\delta(\lambda_{R,j}))-\frac{N_S\lambda_{R,j}\delta(\lambda_{R,j})}{1+\lambda_{R,j}\delta(\lambda_{R,j})}.
	\end{split}
\end{equation}

	Then, by utilizing the property of the unitary matrix $\mathbf{U}_T$, \eqref{averrhoj1} and \eqref{deltadef1} are equivalent to 
\eqref{averrhoj0} and \eqref{fixeq1}, respectively, which completes the proof.

	\section{Proof of Proposition \ref{TheoIlb}}
	\label{proofTheoIlb}
	
	Since $\log (1+x)-\frac{x}{1+x} \geq 0$ for $x\geq0$, we have
	\begin{equation}\label{varrhojAB}
		\begin{split}	
			&\bar{\varrho}_{j}(\mathbf{F})\geq\log \left\vert\mathbf{I}_{N_T}+\frac{\lambda_{R,j}}{1+\lambda_{R,j}\delta(\lambda_{R,j})} \mathbf{T}(\mathbf{F}) \right\vert \\
			& \overset{(a)}{\geq} \frac{1}{1+\lambda_{R,j}\delta(\lambda_{R,j})}\log \left\vert\mathbf{I}_{N_T}+\lambda_{R,j} \mathbf{T}(\mathbf{F}) \right\vert,
		\end{split}	
	\end{equation}
	where step (a) follows the Jensen's inequality. Recalling \eqref{fixeq1}, we have
	\begin{equation}
		\delta(\rho) = \frac{1}{N_S}\sum_{i=1}^{r_T}   \lambda_{T,i}\left(1+\frac{\rho}{1+\rho\delta(\rho)} \lambda_{T,i}\right)^{-1},
	\end{equation}
where $r_T = \mathrm{rank}(\mathbf{T}) \leq \min\{K,\mathrm{rank}(\mathbf{F}\mathbf{F}^\HH)\}$ and $\lambda_{T,i}$ denotes the $i$th non-zero eigenvalue of $\mathbf{T}$. Given $a>0$,  $\frac{x}{1+ax}$ is monotonically increasing for $x\geq0$. Thus, we have
	\begin{equation}
		\delta(\rho) \leq \lim_{\lambda_{T,i} \to \infty}\delta(\rho)= \frac{r_T(1+\rho\delta(\rho))}{\rho N_S} .
	\end{equation}
	Given  $N_S \geq K \geq r_T$, we have 
	\begin{equation}\label{deltaub}
		\delta(\rho) \leq \frac{r_T}{\rho(N_S-r_T)}. %\frac{1}{\rho}\cdot\frac{r_T}{N_S-r_T}.
	\end{equation}
	By substituting \eqref{deltaub} into \eqref{varrhojAB}, we have
	\begin{equation}\label{varrhojAB2}
		\begin{split}	
			&\bar{\varrho}_{j}(\mathbf{F})\geq \frac{N_S -r_T}{N_S}\log \left\vert\mathbf{I}_{N_T}+\lambda_{R,j} \mathbf{T}(\mathbf{F}) \right\vert.
		\end{split}	
	\end{equation}
	By taking the summation of \eqref{varrhojAB2} over index $j$, \eqref{Ilower bound} can obtained.

		\section{Proof of Proposition \ref{lemmaderiveRs}}
		\label{prooflemmaderiveRs}
		%To simplify the notation, we denote $\delta(\lambda_{R,j})$ as $\delta$ in this proof. 
From \eqref{fixeq1}, we have $\delta(\rho) = \frac{1}{N_S} \tr \mathbf{T}\mathbf{M}_T,$ 
%	\begin{equation}
%		\begin{split}
%		\delta(\rho) = \frac{1}{N_S} \tr \mathbf{T}\mathbf{M}_T,
%	\end{split}
%	\end{equation}
where  $\mathbf{M}_T(\rho)\triangleq\left(\mathbf{I}_{N_S}+\alpha(\rho) \mathbf{T}\right)^{-1}$ and $\alpha(\rho)=\frac{\rho}{1+\rho\delta(\rho)}$. 
%\begin{equation}
%	\begin{split}
%\mathbf{M}_T(\rho)\triangleq\left(\mathbf{I}_{N_S}+\alpha(\rho) \mathbf{T}\right)^{-1},
%	\end{split}
%\end{equation}
%and 
%\begin{equation}
%	\begin{split}
%		\alpha(\rho)=\frac{\rho}{1+\rho\delta(\rho)}.
%	\end{split}
%\end{equation}
The derivative of $\alpha(\rho)$ with respect to the $(m,n)$th entry of $\mathbf{F}$, denoted by $F_{m,n}^*$, is given by
\begin{equation}
	\begin{split}
		\alpha_{m,n}'(\rho)&=\frac{\partial \alpha(\rho)}{\partial F_{m,n}^*}=-\alpha^2(\rho) [\mathbf{\Delta}_{\rho}'(\mathbf{F})]_{m,n}.
	\end{split}
\end{equation}
%Note that $\mathbf{T}$ is dependent on $\mathbf{F}$. 
The derivative of $\mathbf{T}$ with respect to $F_{m,n}^*$ can be expressed as
	\begin{equation}
		\begin{split}
			[{\mathbf{T}'}]_{m,n}=\frac{\partial \mathbf{T}}{\partial F_{m,n}^*}= \mathbf{R}_T^{\frac{1}{2}} \mathbf{F}\mathbf{e}_n \mathbf{e}_m^\TT \mathbf{R}_T^{\frac{1}{2}}.
		\end{split}
	\end{equation}
	Therefore, we have
\begin{equation}\nonumber
	\begin{split}
	&[\mathbf{\Delta}_{\rho}'(\mathbf{F})]_{m,n}=\frac{\partial \frac{1}{N_S}\tr\;\mathbf{T}\left(\mathbf{I}_{N_S}+\alpha(\rho)\mathbf{T}\right)^{-1}}{\partial F_{m,n}^*}\\
	&=\frac{1}{N_S}\tr\; [{\mathbf{T}'}]_{m,n} \mathbf{M}_{T}(\rho) \\
	&\quad - \frac{1}{N_S}\tr\; \mathbf{T} \mathbf{M}_{T}(\rho)
	\left(\alpha_{m,n}'(\rho) \mathbf{T} + \alpha(\rho) [{\mathbf{T}'}]_{m,n} \right)\mathbf{M}_{T}(\rho)\\
	%&= \frac{\alpha^2(\rho)\tr\left(\mathbf{T} \mathbf{M}_{T}(\rho)\right)^2 }{N_S} [\mathbf{\Delta}_{\rho}'(\mathbf{F})]_{m,n}\\
	%&\quad +\frac{1}{N_S}\tr \left[ \left(\mathbf{I}-\alpha(\rho)\mathbf{T}\mathbf{M}_T(\rho)\right)[{\mathbf{T}'}]_{m,n} \mathbf{M}_{T}(\rho)\right]\\
	&=  \frac{\alpha^2(\rho)\tr\left(\mathbf{T} \mathbf{M}_{T}(\rho)\right)^2 }{N_S} [\mathbf{\Delta}_{\rho}'(\mathbf{F})]_{m,n}+\frac{[\mathbf{R}_T^{\frac{1}{2}}\mathbf{M}_T^2(\rho)\mathbf{R}_T^{\frac{1}{2}}\mathbf{F}]_{m,n}}{N_S}.
	\end{split}
\end{equation}
		The solution to the above linear equation is given by 
			\begin{equation}\label{derivaLER2}
					\begin{split}
								[\mathbf{\Delta}_{\rho}'(\mathbf{F})]_{m,n} = \frac{[\mathbf{R}_T^{\frac{1}{2}}\mathbf{M}_T^2(\rho)\mathbf{R}_T^{\frac{1}{2}}\mathbf{F}]_{m,n}}{N_S-\alpha^2(\rho)\tr\left(\mathbf{T}(\mathbf{F}) \mathbf{M}_{T}(\rho)\right)^2 }.
						\end{split}
				\end{equation}
%			where $\mathbf{M}_{T}(\rho) =\left(\mathbf{I}+\frac{\rho}{1+\rho\delta(\rho)} \mathbf{T}(\mathbf{F})\right)^{-1}$.
%			\begin{equation}\label{derivaLER}
%				\begin{split}
%						&[\mathbf{\Delta}_{\rho}'(\mathbf{F})]_{m,n} \\
%						&= \frac{\left[\mathbf{R}_T^{\frac{1}{2}}\left(\mathbf{I}+\frac{\rho}{1+\rho\delta(\rho)} \mathbf{T}(\mathbf{F})\right)^{-2}\mathbf{R}_T^{\frac{1}{2}}\mathbf{F}\right]_{m,n}}{N_S-\alpha^2(\rho)\tr\left(\mathbf{T}(\mathbf{F}) \left(\mathbf{I}+\frac{\rho}{1+\rho\delta(\rho)} \mathbf{T}(\mathbf{F})\right)^{-1}\right)^2 }.
%					\end{split}
%			\end{equation}
By consolidating all $[\mathbf{\Delta}_{\rho}'(\mathbf{F})]_{m,n}$ into one matrix, we can obtain \eqref{derivaLER}.

\end{document}